\documentclass[runningheads]{llncs}
\usepackage{graphicx}
\usepackage{booktabs}
\begin{document}
\title{An Interpretable Machine Learning Model with Deep Learning-based Imaging Biomarkers for Diagnosis of Alzheimer’s Disease}
\titlerunning{Interpretable Machine Learning for Diagnosis of Alzheimer’s Disease}
%


\author{Wenjie Kang\thanks{w.kang@erasmusmc.nl}\inst{1}\ \and
Bo Li\inst{1}\ \and
Janne M. Papma\inst{2}\ \and
Lize C. Jiskoot\inst{2}\ \and
Peter Paul De Deyn\inst{3}\ \and
Geert Jan Biessels\inst{4}\ \and
Jurgen A.H. R. Claassen\inst{5}\ \and
Huub A.M. Middelkoop\inst{6,7}\ \and
Wiesje M. van der Flier\inst{8}\ \and
Inez H.G.B. Ramakers\inst{9}\ \and
Stefan Klein\inst{1}\ \and
Esther E. Bron\inst{1}\ \and
for the Alzheimer’s Disease Neuroimaging Initiative\ \and
on behalf of the Parelsnoer
Neurodegenerative Diseases study group}

\authorrunning{W. Kang et al.}
%
\institute{Department of Radiology \& Nuclear Medicine, Erasmus MC, Rotterdam, The Netherlands\ \and
Department of Neurology, Erasmus MC, Rotterdam, The Netherlands\ \and
Department of Neurology \& Alzheimer Center, University Medical Center Groningen, Groningen, The Netherlands\ \and
Department of Neurology, UMC Utrecht Brain Center, University Medical Center Utrecht, Utrecht, The Netherlands\ \and
Radboud University Medical Center, Nijmegen, The Netherlands\ \and
Department of Neurology \& Neuropsychology, Leiden University Medical Center, Leiden, The Netherlands\ \and
Institute of Psychology, Health, Medical and Neuropsychology Unit, Leiden University, The Netherlands\ \and
Amsterdam University Medical Center, location VUmc, Amsterdam, The Netherlands\ \and
Alzheimer Center Limburg, School for Mental Health and Neuroscience (MHeNS), Maastricht University Medical Center, Maastricht, The Netherlands}

%
\maketitle           

\begin{abstract}
Machine learning methods have shown large potential for the automatic early diagnosis of Alzheimer's Disease (AD). However, some machine learning methods based on imaging data have poor interpretability because it is usually unclear how they make their decisions. Explainable Boosting Machines (EBMs) are interpretable machine learning models based on the statistical framework of generalized additive modeling, but have so far only been used for tabular data. Therefore, we propose a framework that combines the strength of EBM with high-dimensional imaging data using deep learning-based feature extraction. The proposed framework is interpretable because it provides the importance of each feature. We validated the proposed framework on the Alzheimer’s Disease Neuroimaging Initiative (ADNI) dataset, achieving accuracy of 0.883 and area-under-the-curve (AUC) of 0.970 on AD and control classification. Furthermore, we validated the proposed framework on an external testing set, achieving accuracy of 0.778 and AUC of 0.887 on AD and subjective cognitive decline (SCD) classification. The proposed framework significantly outperformed an EBM model using volume biomarkers instead of deep learning-based features, as well as an end-to-end convolutional neural network (CNN) with optimized architecture.
\end{abstract}
Code availability:\par
To be added to: https://gitlab.com/radiology/neuro/wenjie-project
\section{Introduction}
Dementia is a major global health problem \cite{prince2015world}. However, early and accurate diagnosis of AD (Alzheimer’s Disease) is challenging \cite{van2013time}. Machine learning methods have shown large potential for early detection and prediction of AD because they can learn subtle patterns and capture slight tissue alterations in high-dimensional imaging data \cite{bron2015standardized,wen2020convolutional}. Nevertheless, those machine learning methods with high diagnostic performance, such as deep learning, are considered black boxes because of the poor interpretability of the predicted results \cite{arrieta2020explainable}.  On the other hand, intrinsically interpretable methods can provide explainable results but often have worse predictive performance as they cannot fully exploit the high-dimensional data \cite{ahmad2018interpretable}. To this end, to facilitate the translation of machine learning to clinical practice it is crucial to find an optimal tradeoff between the accuracy and interpretability.\par 
To solve this problem, we built an interpretable machine learning framework that meanwhile makes use of high-dimensional imaging features. Recently, Explainable Boosting Machines (EBMs) \cite{lou2012intelligible} is a tree-based Generalized Additive Model (GAM) \cite{hastie2017generalized} which have shown comparable accuracy to the state-of-the-art conventional machine learning methods \cite{nori2019interpretml}, and meanwhile provide the contribution to the final decision by each feature for interpretability. Currently, EBM takes as input only tabular data and have not been used with imaging data. By exploit high-dimensional biomarkers from imaging data, we expect the combination of EBM and deep learning techniques will contribute to a more interpretable and accurate prediction for AD diagnosis.\par 

In this work, we propose a framework to extract high-dimensional features from brain MRIs, i.e., deep learning-based imaging biomarkers (DL-biomarkers), which is used by EBM for AD diagnosis. We designed a data-driven strategy to select the region of interest (ROI). Convolutional Neural Networks (CNNs) \cite{krizhevsky2017imagenet} were used to extract DL-biomarkers from whole brain MRI and ROIs. EBM trained with DL-biomarkers can give the importance of each DL-biomarker. For model validation, we compared the performance of the proposed model with an EBM that uses the volumes of brain regions (V-biomarkers) as input. In addition, we compared the proposed model with a CNN with optimized architecture to investigate whether the proposed model maintains comparable diagnostic performance to black-box models. The validation was conducted on both the publicly accessible ADNI dataset, and an external test set.

\section{Methods}
\subsection{Study population}
We used data from two studies. The first group of 855 participants was included from the Alzheimer’s Disease Neuroimaging Initiative (ADNI). We included participants with T1-weighted (T1w) MRI scans available at the baseline timepoint from the ADNI1/GO/2 cohorts, consisting of 335 AD patients, and 520 control participants (CN). The CN group consisted of 414 cognitively normal participants and 106 participants with subjective cognitive decline (SCD). The second group of 336 participants was included from the Health-RI Parelsnoer Neurodegenerative Diseases Biobank (PND) \cite{aalten2014dutch}, a collaborative biobanking initiative of the eight university medical centers in the Netherlands. We included participants at baseline timepoints, including 198 AD patients, and 138 participants with SCD.
\subsection{Data preprocessing} 
The T1w scans were preprocessed following the same pipeline as in \cite{bron2021cross}. After the construction of a dataset-specific template, we computed probabilistic gray matter (GM) maps with the unified tissue segmentation method from SPM8 \cite{ashburner2005unified}. Thereafter, the pre-processed GM maps were cropped to 150 × 180 × 150 voxels to remove the background region.
\subsection{Explainable Boosting Machine (EBM)} 
EBM is a subclass of GAMs which based on trees \cite{chang2021interpretable}. Given a dataset $D$ = \{$(x_i$, $y_i)$\}$_1^N$, where for any subject $i\in[1, N]$, $x_{\mbox{\scriptsize i}}$ = $(x_{i1}$, ..., $x_{in})$ is a feature vector with $n$ features, and $y_i$ is the label, EBM is of the form:\par
$$
g(Y)=\beta + \sum f_j\left(X_j\right)+\sum f_{ij}\left(X_i, X_j\right),
$$
where $g(.)$ is the link function that adapts the classification setting, $X_j$ is the $j$th feature vector in the feature space, $Y$ is the target class, and shape functions $f_j$ and $f_{ij}$  ($i \ne j$) are gradient-boosted ensembles of bagged trees in EBM. As a subclass of GAM, EBMs prevent interaction effects from being learned. The ability to analyze features independently makes EBMs easy to reason about the contribution of each feature to the prediction \cite{caruana2015intelligible}. EBMs include pairwise interaction terms $f_{ij}$$(X_i$, $X_j)$ \cite{lou2013accurate}, it finds all the possible pairs of interactions in the residuals and
orders them by importance. A previous work using EBM have applied it to volumetrics and yielded an AUC of 0.842 in the task of predicting whether mild cognitive impaired (MCI) patients will convert to AD \cite{sarica2021explainable}. However, such features of regional volume are a crude summary of the high-resolution brain images, only part of the information presented in the images is included in the regional volume.

\subsection{Proposed extension}
 Here, we propose to use deep learning models to extract features (DL-biomarkers) from high-dimensional brain MRIs. The predicted results of CNNs for binary classification which are the probability of positive are used as the DL-biomarkers. For the full brain DL-biomarker, a CNN is trained in classification task that takes whole-brain images as input (Global CNN, Glo-CNN). For the regional DL-biomakers, lightweight CNNs are trained that take selected image patches as input (Local CNN, Loc-CNN). The architecture of our proposed EBM is shown in Fig 1. Each input image is handled by a shape function which is a trained CNN.
\begin{figure}
\centering
\includegraphics[width=0.75\textwidth]{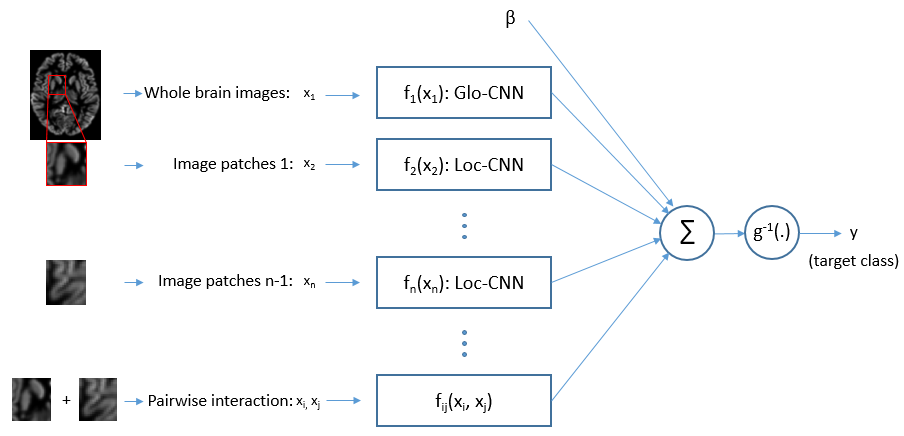}
\caption{Architecture of the proposed EBM using DL-biomarkers. Shape function f$_{1-n}$ outputs a DL-biomarker for the whole brain or a brain region and f$_{ij}$ outputs a pairwise DL-biomarker. $\Sigma$ is the weighted sum of shape functions and $g^{-1}(.)$ is the activation function.} \label{fig1}
\end{figure}

\subsection{Extraction of the DL-biomarkers}
The architectures of the Glo-CNN and Loc-CNN are adapted from the state-of-the-art research \cite{cui2019rnn,dyrba2021improving}. We provide the details of the CNNs optimized for AD diagnosis in the supplement. The DL-biomarkers of the subjects in the testing set were predicted by trained CNNs. Loc-CNNs shared the same architecture but trained on different image patches. To select patches for brain regional features, we used an occlusion map strategy \cite{zeiler2014visualizing,rieke2018visualizing}. This method computes the impact on the output of the network by occluding a patch in the input image. The gap between the two outputs, therefore, shows quantitatively how much an image patch affects the decision-making of the deep learning model. We compute subject-level occlusion maps by sliding an occlusion patch across the whole image. Subject-level occlusion maps in testing set were summed into a group-level map, which was used for patch selection. Within the group-level occlusion map, total weight per patch was computed, and patches were ranked and selected according to the total weight. With the selected patches, Loc-CNNs were trained to predict the regional DL-biomarkers. The value of the DL-biomarkers indicates the probability of the subject being positive. We provide a flowchart of the extraction of DL-biomarkers in Fig. 2, and implementation details in section \ref{ss3.2}.
\begin{figure}
\centering
\includegraphics[width=0.8\textwidth]{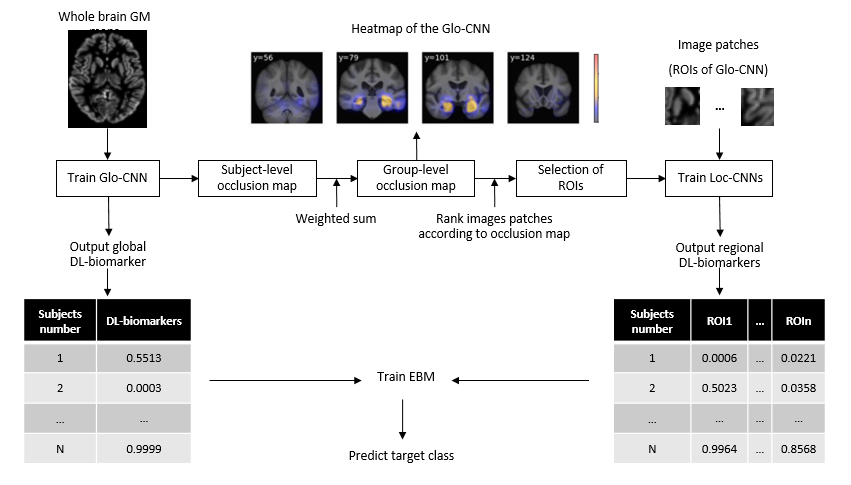}
\caption{The flowchart of the extraction of DL-biomarkers.} \label{fig2}
\end{figure}

\section{Experiments} 
\subsection{Validation study} 
We compared the proposed EBM using DL-biomarkers with two baseline methods: a CNN and an EBM using V-biomarkers, for AD-CN classification. For the validation on ADNI, AD and CN groups were split in a stratified way into an optimization set and a clean testing set (test$_{split}$) in a ratio of 9 : 1. Glo-CNNs and Loc-CNNs were trained in 5-fold cross-validation in the optimization set. Occlusion maps and imaging biomarkers for all the subjects on the testing set of each fold (test$_{cv}$) were gathered. The models trained on ADNI were also validated on external testing set PND.\par
We used accuracy (ACC), sensitivity (SEN), specificity (SPE), and area-under-the-curve (AUC) as the performance metrics for binary classification. We validated the performance of the proposed CNN model separately. To test the performance of Glo-CNN on ADNI, AD and CN groups were randomly split for 10 iterations preserving relative class sizes, using 90\% for training and 10\% for testing. Confidence intervals (95\%CIs) for the mean performance measures were calculated using the corrected resampled t-test \cite{nadeau1999inference}. For the validation of Glo-CNN on PND, we used all participants on ADNI as the training set and used PND as the external testing set. 95\%CIs were obtained based on 100 repetitions of bootstrap on testing set. For the validation of the proposed EBM and the two baseline methods, we trained models on the split optimization set on ADNI, and test the methods on the test$_{split}$ and PND. 95\%CIs were obtained based on 10 repetitions of bootstrap on the testing set.

\subsection{EBM using DL-biomarkers} \label{ss3.2}

The Glo-CNN and Loc-CNN were compiled with a class balanced binary cross-entropy loss function. All CNN models used Adam optimizer \cite{kingma2014adam}. The initial leaning rate was 5×10$^{-4}$, and models were trained using the batch size of 16. The group-level occlusion maps were obtained by the summation of the subject-levavel occlusion maps obtained from subjects in (test$_{cv}$) using trained Glo-CNNs. The occlusion patch has a size of $20^3$. Occlution maps resized to the same size of the original images using bicubic interpolation. We chose the top 10 ROIs with a patch size of $30^3$ based on the group-level occlusion map, and trained ten Loc-CNNs based on each of the ten ROIs to predict regional DL-biomarkers.

\subsection{Baseline methods} 
We used Glo-CNN as the baseline CNN model, because it was optimized for the best performance in the validation set among all CNN models. For the baseline EBM, the GM volume of each brain region is considered as a volume biomarker (V-biomarkers). The GM volumes were corrected by intra-cranial volume \cite{gousias2008automatic,hammers2003three}. V-biomarkers were named after brain regions, with `Total brain' indicating the GM volume of the whole brain. We chose the top 11 among 75 V-biomarkers with the smallest p-values between AD and CN groups in the optimization set. The EBMs based on DL-biomarkers and V-biomarkers both used 11 biomarkers, and also included top-2 pairwise biomarkers. In addition, we took the average output of the Glo-CNN and Loc-CNNs as the output of a CNN ensemble model (Glo/Loc-CNN) which has the similar computational complexity as the EBM based on DL-biomarkers. We provide the performance of the Glo/Loc-CNN in the supplement.

\section{Results} 
\subsection{Glo-CNN results} 
We provide the details of the cross-validation results of Glo-CNN in the supplement. The model yielded an ACC of 0.880 (95\%CI: 0.852-0.908) and an AUC of 0.944 (0.870-1.00) on the AD-CN task on ADNI. On the external PND test set, the model yielded an ACC of 0.789 (0.744-0.834) and an AUC of 0.872 (0.830-0.914) on the AD-SCD task.

\subsection{Occlusion map and ROIs}
The selected 10 ROIs based on the Glo-CNN are shown in Fig. 3 (a), the overlap of ROIs is demonstrated by colors. We name the selected ROIs after the overlapped brain regions. The location and name of ROIs in the coronal plane are shown in Fig. 3 (b).

\begin{figure}
\centering
\includegraphics[width=0.9\textwidth]{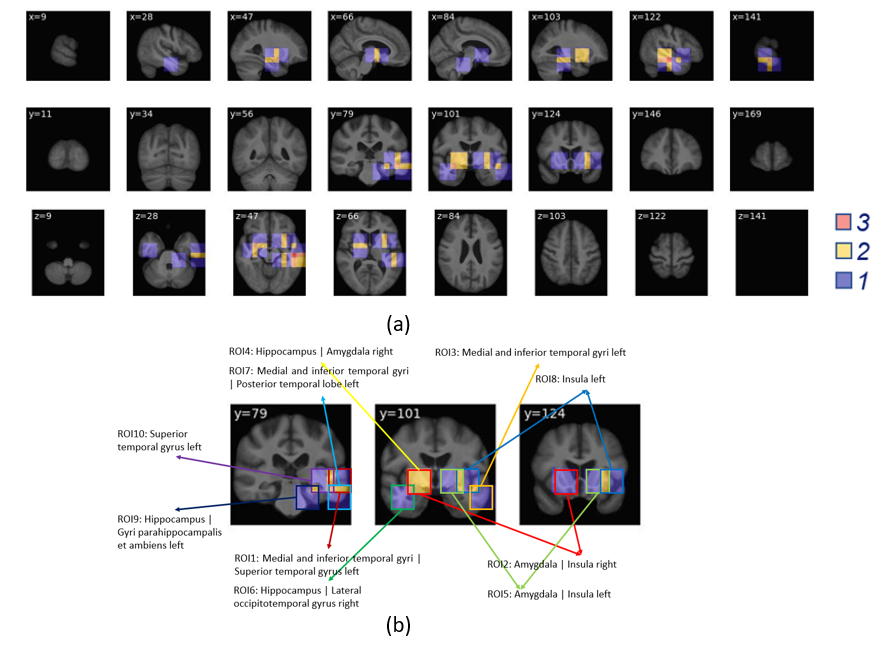}
\caption{(a) The location, and (b) the ID and name of the selected image patches.} \label{fig4}
\end{figure}

\subsection{Comparison study}
The performance of the methods on the ADNI dataset is shown in Fig. 4 (a). The accuracy (0.883) and AUC (0.970) of the proposed EBM using DL-biomarkers were significantly higher (p-value \textless 0.01) than those of the EBM using V-biomarkers (ACC=0.815; AUC=0.899). The AUC (0.970) of EBM using DL-biomarkers was significantly higher than Glo-CNN (0.944), while the accuracy (0.883) was not significantly higher than Glo-CNN (ACC=0.873, p-value = 0.08). The performance of the methods on the external testing set PND is shown in Fig. 4 (b). The accuracy (0.778) and AUC (0.887) of EBM trained with DL-biomarkers were significantly higher than those of the EBM trained with V-biomarkers (ACC=0.703; AUC=0.853). The AUC (0.887) of EBM trained with DL-biomarkers was significantly higher than Glo-CNN (0.852), while the accuracy did not have significant difference (p-value = 0.4). The group-level feature importance of EBMs trained with DL-biomarkers and V-biomarkers in training set are reported in Fig. 5. The two pairwise biomarkers in the EBM trained with DL-biomarkers are the combination of Total brain with ROI1, and the combination of Total brain with ROI6. The feature importance shown in Fig. 5 can show which brain regions highly affect the decision-making of EBMs.

\begin{figure}
\centering
\includegraphics[width=0.8\textwidth]{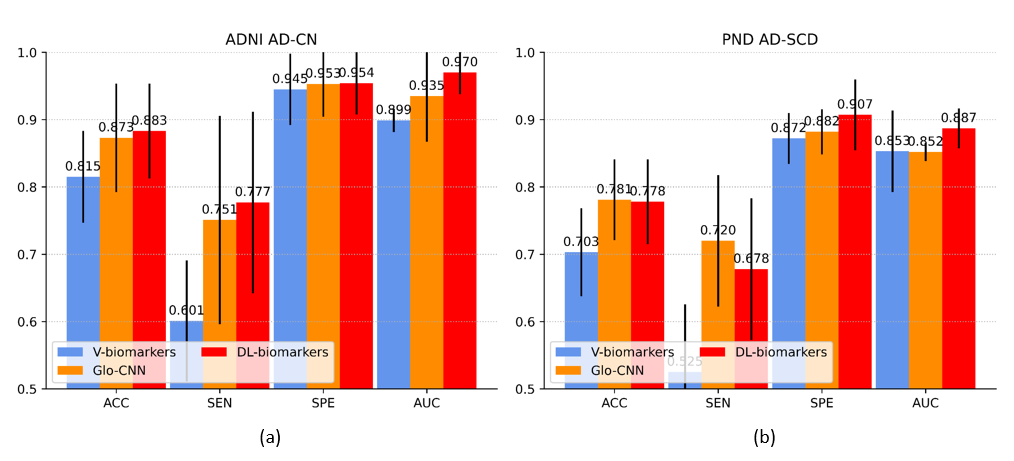}
\caption{The performance of the EBM trained with V-biomarkers (V-biomarkers; blue), Glo-CNN as the CNN baseline (Glo-CNN; orange), the EBM trained with DL-biomarkers (DL-biomarkers; red) (a) on ADNI (b) on external testing set PND. The error bars represent for the confidence intervals.} \label{fig5}
\end{figure}

\begin{figure}
\includegraphics[width=\textwidth]{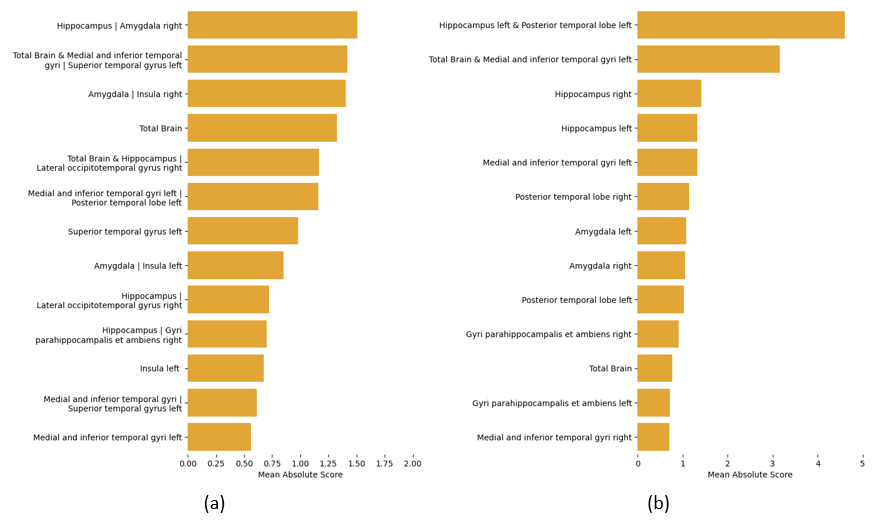}
\caption{The feature importance of (a) the proposed EBM using DL-biomarkers, and (b) EBM trained with V-biomarkers.} \label{fig6}
\end{figure}

\section{Discussion and Conclusions}
In this paper, we proposed a new EBM using high-dimensional imaging biomarkers predicted by CNNs for AD diagnosis, which used occlusion maps for region selection. The proposed framework is more interpretable than black-box models because it provides the feature importance of all biomarkers. We compared the performances among the proposed EBM using DL-biomarkers, the standard EBM using V-biomarkers, and a CNN baseline. The results show that our proposed framework yielded higher classification performance than the other two methods in the AD-CN task on the ADNI dataset. We suspect that the improved performance is because DL-biomarkers are the predictions of Glo-CNN and Loc-CNNs, which enables the encoding of high-dimensional features \cite{bo2023}. Furthermore, the results of the EBM trained with DL-biomarkers yielded higher AUC than the CNN model. We assume this is because EBM is an ensemble of models that extract complementary features from brain regions. However, the performance on the PND dataset was lower than that on the ADNI, we assume that it is because the distributions of data differ between the two cohorts and the classification task differs from AD-CN to AD-SCD.\par

In conclusion, our proposed method achieved higher classification performance than the baseline models and allows for the interpretation of the brain features relevant to AD diagnosis based on imaging data. In future work, we intend to further explore the interpretability of the proposed method. We also intend to validate the generalization of our method. Further validation will include dementia prediction in MCI and diagnosis of other subtypes of dementia.

\newpage
%
%
%

\end{document}